\documentclass[prb,amsmath,amssymb,showpacs,twocolumn]{revtex4}

\pdfoutput=1
\usepackage{graphicx}
\usepackage{dcolumn}
\usepackage{bm}
\usepackage{hyperref}
\usepackage{amsmath}
\usepackage[utf8]{inputenc}
\usepackage[swedish]{babel}

\begin{document}

\title{Signatures of Coherent Electronic Quasiparticles in the Paramagnetic Mott Insulator.}
\author{Mats Granath$^1$ and Johan Sch\"{o}tt$^2$}
\affiliation{$^1$Department of Physics, University of Gothenburg,
SE-41296 Gothenburg, Sweden}
\affiliation{$^2$Department of Physics and Astronomy, Uppsala University, P.O. Box 516, SE-751 20 Uppsala, Sweden}

\date{\today}

\begin{abstract}
We study the Mott insulating state of the half-filled paramagnetic
Hubbard model within dynamical mean field theory using a recently formulated stochastic and non-perturbative quantum impurity
solver. The method is 
based on calculating the impurity self energy as a sample
average over a representative distribution of impurity models solved
by exact diagonalization. Due to the natural parallelization of the method, millions of
poles are readily generated for the self energy which allows to work
with very small pole-broadening $\eta$. Solutions at small and large
$\eta$ are qualitatively different; solutions at large $\eta$ show
featureless Hubbard bands whereas solutions at small $\eta\leq 0.001$ (in units of half bare band width)
show a band of electronic quasiparticles with very small quasiparticle weight at the inner edge of the Hubbard
bands.  The validity of the results are supported by agreement within
statistical error $\sigma_{\text{QMC}}\sim 10^{-4}$ on the imaginary frequency axis with calculations using a continuous time quantum Monte Carlo
solver. Nevertheless, convergence with respect to finite size of the
stochastic exact diagonalization solver remains
to be rigourously established. 

\end{abstract}


\maketitle

The concept of electronic quasiparticles is one of the most basic
paradigms in the description of the dense interacting electron system
found in any metallic atomic crystal. In terms of quasiparticles even
superficially very strongly interacting systems may be described
through elementary excitations that are in direct
correspondence with those of a non-interacting electron gas. This allows
for a relatively simple description of thermodynamic and transport
properties, as well as the inclusion of interactions between
quasiparticles and phonons which is the basis of the standard
theory of superconductivity as a condensate of paired quasiparticles.

The basic theory of quasiparticles is well
understood within the framework of quantum many particle physics\cite{NegeleOrland} where
the distribution of single electron excitations at (crystal) momentum $\vec{k}$ and energy
$\omega$ ($\hbar=1$) are described by the spectral function
$A_{\vec{k}}(\omega)$. The object that captures the effects of
interactions is the self energy $\Sigma_{\vec{k}}(\omega)$ in terms of
which the spectral function can be expressed as 
$A_{\vec{k}}(\omega)=-\frac{1}{\pi}\frac{Im
  \Sigma_{\vec{k}}(\omega)}{(\omega-\epsilon_{\vec{k}}+\mu-Re\Sigma_{\vec{k}}(\omega))^2+(Im\Sigma_{\vec{k}}(\omega))^2}
$. Here $\epsilon_{\vec{k}}$ is the bare band energy of the system and
$\mu$ is the chemical potential and we have assumed a single band which
is isolated from any other effects than those of intraband
electron-electron interactions.    
In a Fermi liquid the quasiparticles
are objects that close to the Fermi energy,  at energy $\omega=E_{\vec{k}}$ solves
$\omega-\epsilon_{\vec{k}}+\mu-Re\Sigma_{\vec{k}}(\omega)=0$.
These are characterized by a lifetime
  $\tau^{-1}=-2Im\Sigma_{\vec{k}}(\omega=E_{\vec{k}})  $, quasiparticle
  weight $Z_{\vec{k}}^{-1}=1-\partial_\omega
  Re\Sigma_{\vec{k}}(\omega)|_{E_{\vec{k}}}$, and effective mass. The
  quasiparticle weight $Z_{\vec{k}}$ quantifies how much of the spectral weight
  at momentum $\vec{k}$ is carried by the Landau quasiparticle and
 will be manifest as
 the characteristic discontinuity at $T=0$ in the occupation number
 $n_{\vec{k}}=\int_{-\infty}^0d\omega A_{\vec{k}}(\omega)$ which
 defines the Fermi surface.  

For an interacting electron system it can be understood in
perturbation theory how the quasiparticles are a consequence of
phase space constraints for scattering which gives a vanishing
$Im\Sigma (\omega=0)=0$ and a correspondingly divergent lifetime of
quasiparticles at the Fermi energy.\cite{NegeleOrland}

In contrast a Mott insulator is a system
that from the band structure considerations would be a metal but which
has been driven insulating by electron-electron
interactions.\cite{Mott} The defining feature of the insulator is exactly that
there are no low energy electronic quasiparticles, instead there is a
gap in the spectral function around the Fermi
energy.
 In its cleanest form (without broken translational symmetry) the Mott insulator is a non-Fermi liquid in the sense
 that the self energy diverges at small $\omega$, and electronic
 quasiparticles are thus always unexpected.  Quasiparticles, if found,
 would be of fundamental importance and might also provide a setting 
for theories of  more exotic states in strongly correlated systems.

In this paper we study the Mott insulator in infinite dimensions within dynamical mean field
theory (DMFT) \cite{George_review,DMFT1,DMFT2,DMFT,DMFT3}, for the standard case of the single band paramagnetic
Hubbard model with a semi-circular density of states. Although this is
a much studied\cite{IPT,ED,Local_moment,NRG,strong_coupling,Nishimoto,DMRG_Garcia,Karski_DMRG,Dist-ED,Lu} and by now perhaps the archetypical model of a Mott insulator
(and Mott transition), we a found evidence that a basic feature of the
spectral function may be missing in most earlier 
studies; namely that it may contain well defined electronic
quasiparticles at the gap edge. We have used
a stochastic non-perturbative method, ''distributional exact
diagonalization'' (Dist-ED)\cite{Dist-ED} which can give exceptionally
high resolution of the self energy over the full band width. The
method in brief consists of calculating the self energy of the DMFT
(Anderson) impurity problem as a sample average over representative but
stochastically generated finite size impurity models that are solved
by exact diagonalization. In standard fashion a finite shift $\eta > 0$ away
from the real axis is used in order to generate a continuous function
from a finite set of poles, but due to the natural parallelization of the present method, 
millions or even billions of poles can be generated which allows for the use of very
small $\eta$. Quasiparticles are found only at $\eta\leq
0.001$ (in units of half bare band width) and for not too large $U$, suggesting that both of
these aspects, non-perturbative and high resolution, are crucial to
the result. In order to see this feature it is thus necessary to be able solve the DMFT
equations non-perturbatively and very close to the real axis which is difficult with
other methods at the high energies related to the Mott gap.

\begin{figure}
\includegraphics[scale=.5]{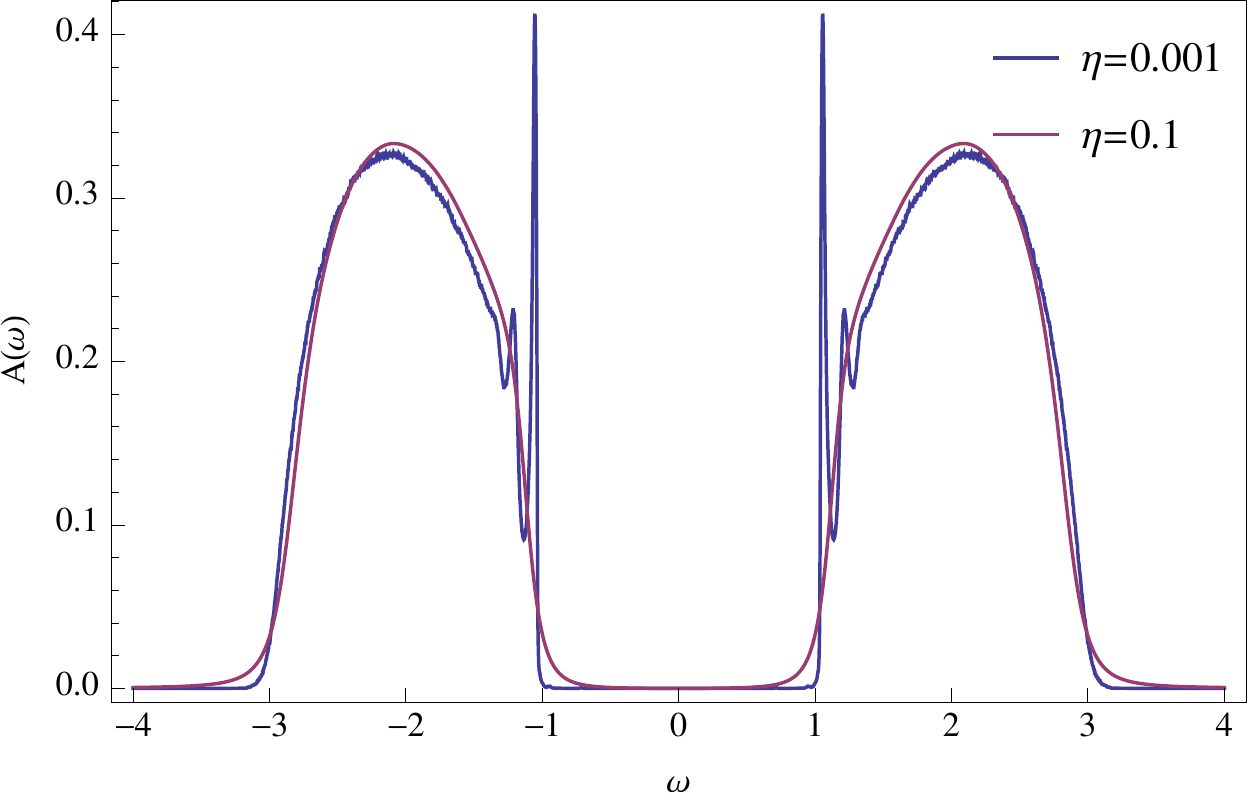}
\includegraphics[scale=.5]{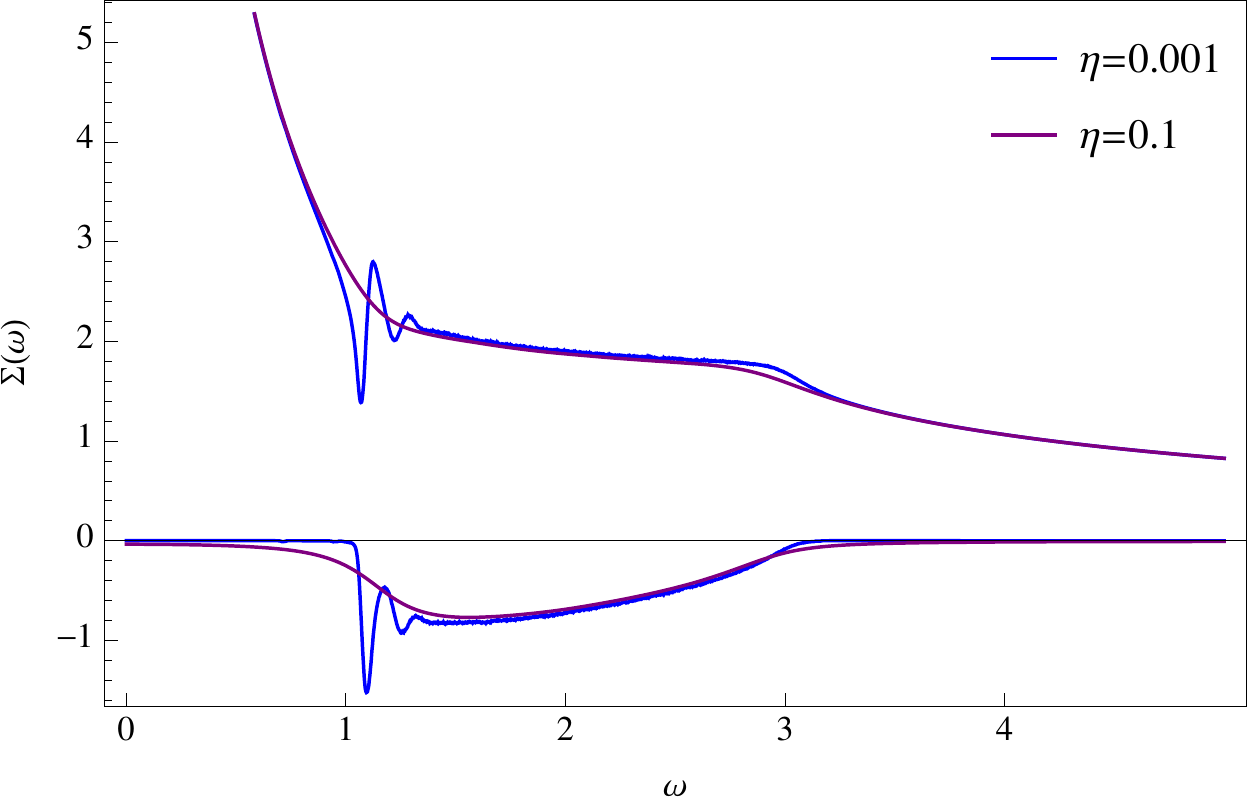}
\caption{\label{delta001_delta01}
Converged solutions at $U=4$ ($n_s=5$) for large ($\eta=0.1$) and small ($\eta=0.001$) broadening, 
showing the upper and lower Hubbard bands
as well as the very sharp peak at the inner edge. (Bottom)
Corresponding real
and imaginary parts of the self energy. (The pole at $\omega=0$ is
not shown in $Im\Sigma$.) The $\eta=0.001$ calculation is an average of $6\cdot
10^6$ samples with a corresponding self energy consisting of $>10^8$ poles.  
}
\end{figure}

\begin{figure}
\includegraphics[scale=.5]{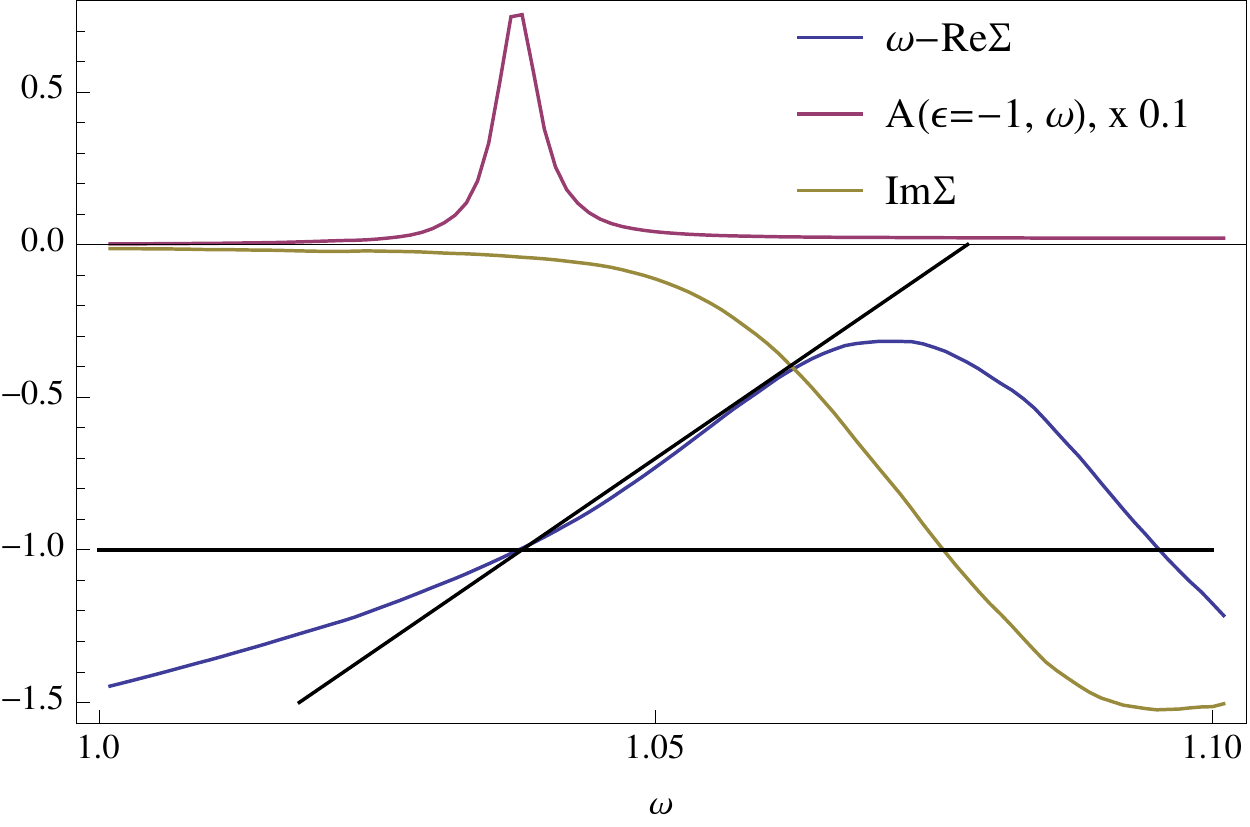}
\caption{\label{sigma_dosU4}
Close-up of the the quasiparticle at $\epsilon=-1$ (not integrated over band energies), together with $Im\Sigma$
and $\omega-Re\Sigma$, showing that this is an actual quasiparticle
that solves $\omega-\epsilon-Re\Sigma=0$ close to the bare band edge
$\epsilon=-1$ with a large lifetime $(Im\Sigma)^{-1}$. The tangent at the crossing correspond to the inverse
quasiparticle weight $Z^{-1}=1-\partial_\omega Re\Sigma\approx 25$. 
}
\end{figure}

\begin{figure}
\includegraphics[scale=.5]{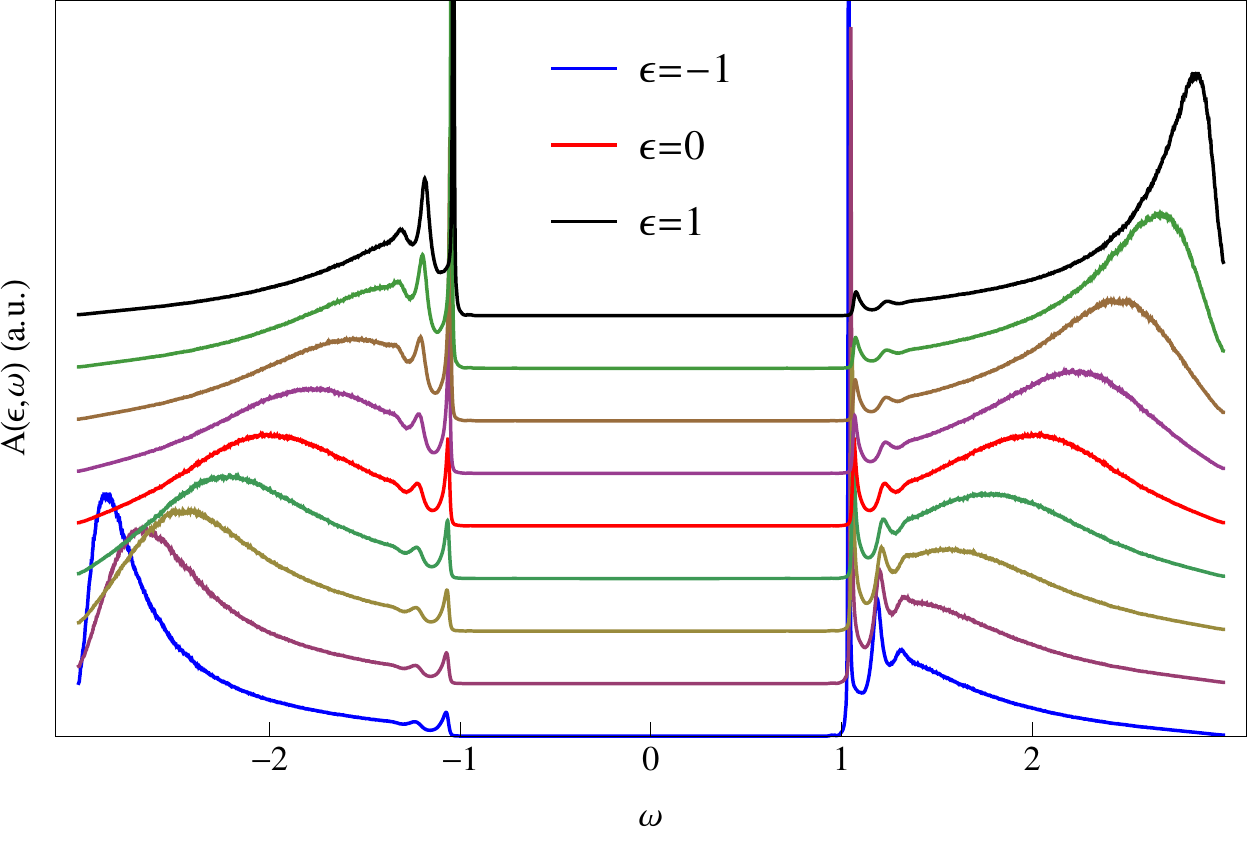}
\caption{\label{dispersion}
Spectral function $A(\epsilon,\omega)$ resolved with respect to band
energy $\epsilon$. Showing the broad incoherent Hubbard bands together
with the very narrow band of gap edge quasiparticles.
}
\end{figure}

\begin{figure}
\includegraphics[scale=.5]{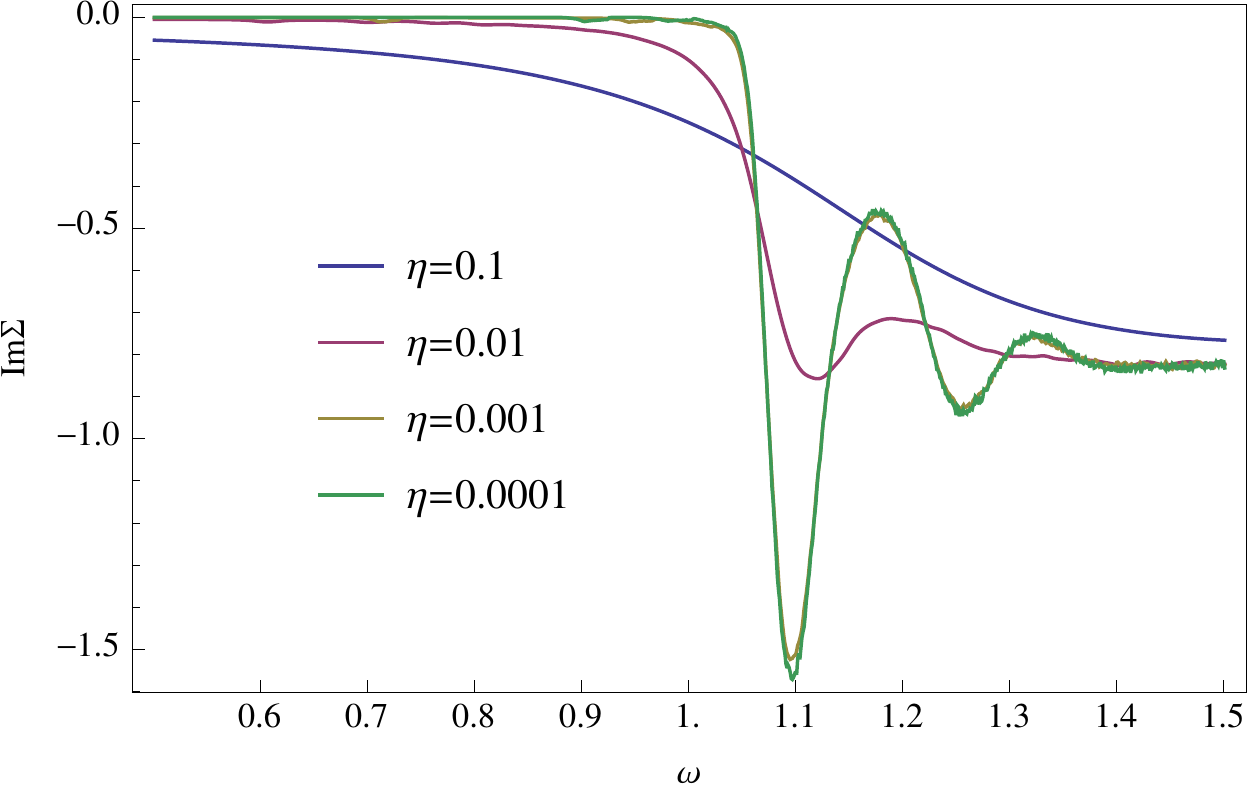}
\caption{\label{sigma_of_delta}
Dependence on $\eta$, broadening of poles, for the imaginary part of
the self-energy, for $\eta=10^{-1},10^{-2},10^{-3},10^{-4}$. To
resolve any peak structure requires a small $\eta\leq
10^{-2}$, with convergence for $\eta\leq 10^{-3}$.
}
\end{figure}

\begin{figure}
\includegraphics[scale=.5]{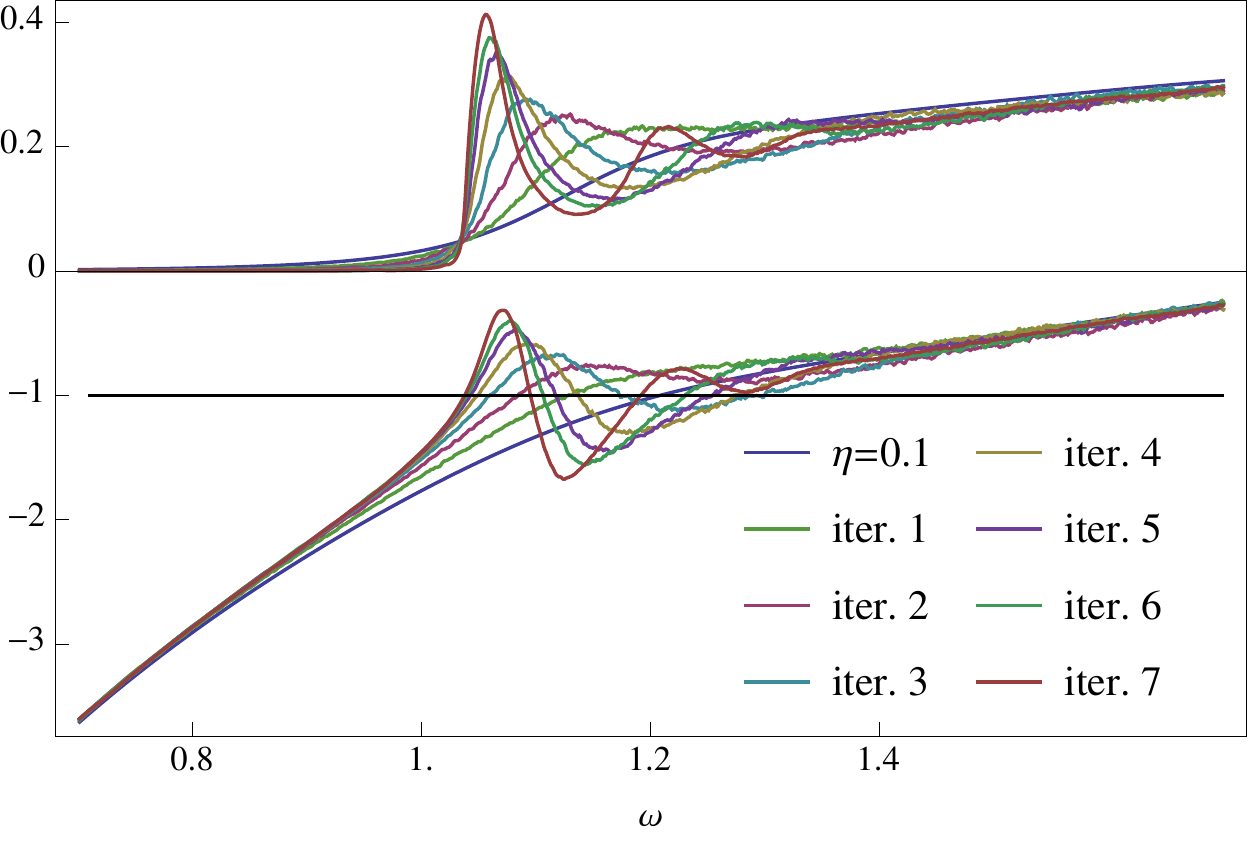}
\caption{\label{DMFT_iteration}
DMFT iterations at $\eta=0.001$, showing $\omega-Re\Sigma(\omega)$ (lower panel) and
corresponding $A(\omega)$, close to inner band edge.  The iterations are started from a (featureless) converged
solution at $\eta=0.1$ and show that the quasiparticle peak is not an
explicit finite size effect but rather a feature that is iteratively enhanced in the DMFT cycle.
}
\end{figure}

\begin{figure}
\includegraphics[scale=.45]{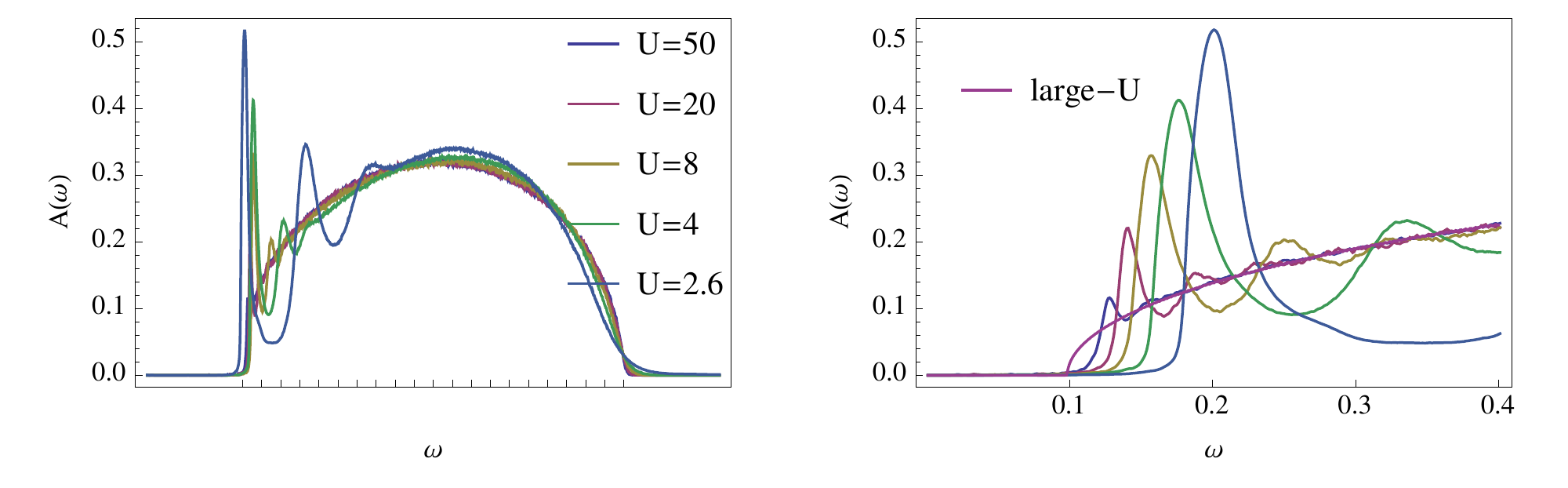}
\caption{\label{various_U}
Upper Hubbard band for various $U=2.6,4,8,20,50$, shifted to same
center (left). 
Inner edge (right) showing the gradual disappearance (arbitrary offset) of the peak structure for large
$U$ and compared to the exact result to lowest order in
$1/U$.\cite{strong_coupling} At the largest $U$ there are no
actual quasiparticles. 
}
\end{figure}

\begin{figure}
\includegraphics[scale=.5]{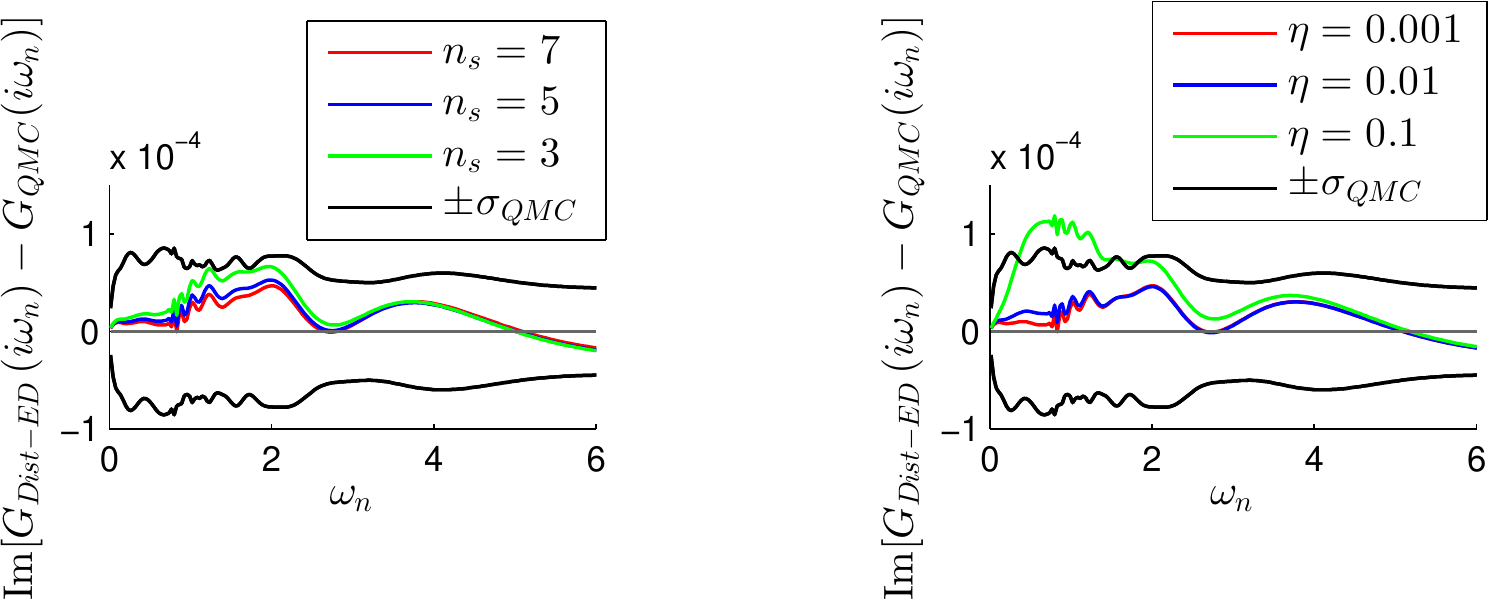}
\caption{\label{QMC_compare}
Dist-ED for various various $n_s$
(using $\eta=0.001$) and various $\eta$ (using $n_s=7$), compared to CT-QMC calculations at inverse temperature
$\beta=200$ and $U=4$ on the Matsubara
frequencies $i\omega_n=\frac{\pi}{\beta}(2n+1)$, with $\sigma_{\text{QMC}}$ the
calculated statistical error of the QMC data.  
}
\end{figure}

The main result of the paper,
as shown in Figure \ref{delta001_delta01},
is the sharp peaks in the local density of states and the imaginary part
of the self energy for small $\eta=0.001$ at the inner edge of the
Hubbard bands, a feature that is not found for larger
$\eta=0.1$. In Figure \ref{sigma_dosU4} it is shown that this edge peak
derives from an actual quasiparticle
solution $\omega-Re\Sigma-\epsilon=0$, for a narrow window of
$\omega$ values and a range of bare
band-energies $\epsilon$. The corresponding  energy resolved
spectral function, $A(\epsilon,\omega)=-\frac{1}{\pi}Im(\frac{1}{w+\mu-\epsilon-\Sigma(\omega)})$,
shows narrow dispersing bands (Fig. \ref{dispersion}) together with
the broad incoherent weight. An important conclusion, as
demonstrated in Figures \ref{sigma_of_delta} and \ref{DMFT_iteration}, is that the quasiparticle
solutions are only observed if the DMFT self consistency cycle is
performed very close to the real axis, and in Figure \ref{delta01_DMRG} that the
small and large $\eta$ solutions are qualitatively
different. This difference can be linked to a small energy scale (see
Figure \ref{e_scale}) of separation between spectral weight,
$A(\omega)$, and scattering, $Im\Sigma(\omega)$, which is a
non-perturbative outcome from exact diagonalization. This energy scale
is diminished at large $U$ such that the known strong coupling form\cite{strong_coupling}
(Fig. \ref{various_U}) of the Hubbard bands is asymptotically
reached. Comparing to calculations using a continuous time quantum
Monte Carlo (CT-QMC) impurity solver\cite{triqs} (Fig. \ref{QMC_compare}) on
imaginary frequencies show that
the small $\eta$ solutions are in agreement within 
statistical error with the QMC calculations.


Similar sharp peaks as found in our study have been reported before in
various forms but are difficult to distinguish from peaks due to
finite resolution of poles. In the metallic phase, there is quite strong
evidence that such gap edge peaks
do appear.\cite{Karski_DMRG,Lu} 
Also, in the antiferromagnetic case such features have been
observed, where they might (at least physically) be understood in terms of
coupling to low energy magnon modes.\cite{Sangiovanni}  
One study, by Nishimoto et al.\cite{Nishimoto}  using dynamical density-matrix
renormalization group (D-DMRG)\cite{DDMRG} finds
additional peak structure even in the insulting paramagnetic phase, although the
details of the self-energy is not explored. As our study is
limited to small system sizes with number of levels $n_s=$3, 5, or 7,
the qualitatively similar results in the D-DMRG study with
$n_s\sim{\cal O}(200)$ is reassuring. Nevertheless, other more recent
studies do not find quasiparticles in the Mott insulating state and it
seems fair to say that this is still an open
question.\cite{Karski_DMRG,Lu,Ganahl}

\subsection*{Model and method}

The model studied is the single band Hubbard model with on-site
interaction quantified by 
$U$ but treated in infinite dimensions where it can be exactly mapped to a quantum impurity
with a momentum independent self energy.\cite{George_review}
We additionally 
assume that there is no broken symmetry such that
there is no magnetic or charge order.  

At half-filling, $\mu=U/2$, and using a semi-circular bare density of states
$\rho_0(\omega)=\frac{2}{\pi}\sqrt{1-\epsilon^2}$ ($|\epsilon|\leq 1$)
of bandwidth 2, the local interacting Greens function (for one spin
species) is
$G(z)=\int d\epsilon\frac{\rho_0(\epsilon)}{z+\mu-\epsilon-\Sigma[G](z)}$, which
can be integrated to the form 
\begin{equation}
G(z)=\frac{1}{z+\mu-\Delta[G](z)-\Sigma[G](z)}\,,
\end{equation}
with $\Delta[G]=(1/4)G$.  Here $\Sigma[G]$ is the self energy of the corresponding Anderson
impurity specified by interaction $U$ and chemical potential $\mu$ on the impurity site, and hybridization between impurity and continuum
$\Delta[G]$. 
 The challenging task is to solve
for the self energy, or equivalently, the interacting Green's function
of the impurity model. 

The standard diagrammatic perturbation theory would involve the
impurity-bath Greens function
$G_0(z)=\frac{1}{z-\Delta}$ as propagators with four point vertex $U$
and two-point vertex $\mu$. (As discussed shortly the inclusion of
$\mu$ in the interaction rather than the propagator is in principle a
matter of convenience.)  
We will not do perturbation theory but nevertheless use $G_0$ (rather
than $\Delta$) to describe the impurity.   
Assuming $\Delta$ is particle-hole symmetric
and with a clean gap we can write 
\begin{equation}
G_0=\frac{a_0}{z}+g_0(z)\,,
\end{equation}
with $a_0=(1-d_\omega\Delta|_{\omega=0})^{-1}$. Here $g_0$ is gapped
such that $Im g_0(\omega+i0^+)=0$ for $|\omega |<\epsilon_{gap}$ and has integrated weight $\int
d\omega\frac{-1}{\pi}Im g_0(w)=1-a_0$. 

The choice of including the chemical potential in the interaction
rather than in the non-interacting Greens function ($G'_0=1/(z+\mu-\Delta)$) is
convenient (and standard for this problem) because it makes $G_0$ particle-hole
symmetric. Within the present approximation the convention is also valuable because
it isolates a substantial part of the spectral weight in a single pole
at $\omega=0$ that may be represented exactly by a finite system, and it also
separates the pole from the continuum with a gap of finite width.


\begin{figure}
\includegraphics[scale=.3]{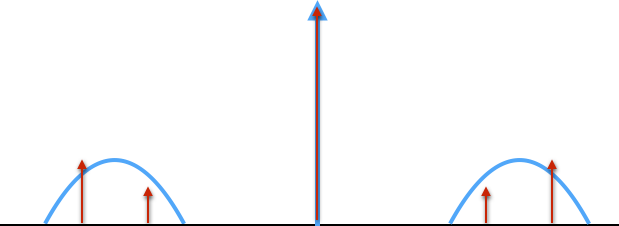}
\caption{\label{example}
Schematic of the sampling of the continuous impurity-bath Greens
function $G_0=\frac{a_0}{z}+g_0(z)$ in terms of a five level ($n_s=5$)
system $G_0^\nu= \frac{a_0}{z}+\sum_{i=1}^{4}\frac{a_i^\nu}{z-b_i^\nu}$. The location of the satelite poles at finite
$\omega$ are generated stochastically based on the distribution $-Im
g_0(\omega)$ with random (normalized) residues.  
}
\end{figure}
As was suggested in Ref. \onlinecite{Dist-ED} and presently specialized to the
{\em gapped} Anderson model we propose to make an approximate calculation of the self energy
in the following way:\\
{\em 1) Generate a large number $N$ of finite-size
Anderson models} (see Fig. \ref{example}) indexed by $\nu=1,...,N$ with $n_s$ orbitals ($n_s$
odd) that correspond to a finite impurity-bath Greens function
\begin{equation}
G_0^{\nu}=\frac{a_0}{z}+\sum_{i=1}^{n_s-1}\frac{a_i^\nu}{z-b_i^\nu}\,.
\end{equation} 
For $i<n_s/2$ the pole
locations $b_i^\nu$ are picked using $ g_0$ as a
probability distribution function, such that 
\begin{equation}
P(b_i^\nu)=-\frac{1}{(1-a_0)\pi}Im\,g_0(\omega=b_i^\nu+i0^+)\,,
\end{equation}
 and $a_i^\nu$ is a
random positive number. Particle-hole symmetry is enforced by taking
$b_i^\nu=-b_{n_s-i}^\nu$ and $a_i^\nu=a_{n_s-i}^\nu$ for $i>n_s/2$,
and normalization by taking $\sum_{i=1}^{n_s-1}a_i^\nu=1-a_0$. The construction ensures
that 
\begin{equation}
\langle G_0^\nu\rangle\equiv\lim\limits_{N\rightarrow\infty}\frac{1}{N}G_0^\nu=G_0\,.
\end{equation}
\\

{\em 2) Identify parameters of the Anderson Hamiltonian} 
\begin{equation}
H^\nu=Un_{0,\uparrow}n_{0,\downarrow}-\mu
n_0-\sum_{i=1,\sigma}^{n_s-1}V_i^\nu(c_{i\sigma}^\dagger
c_{0,\sigma}+\text{h.c.})+\sum_{i=1}^{n_s-1}\epsilon_i^\nu n_{i}
\end{equation}
where $n_{i\sigma}=c_{i\sigma}^\dagger c_{i,\sigma}$ is the number
operator at the bath site $i>0$ or impurity site $i=0$ with
spin $\sigma=\uparrow,\downarrow$ and $n_i$ is summed over
spin. Hopping between impurity and bath site $i$ is given by $V_i^\nu$
and the bath level energy by $\epsilon_i^\nu$.
The non-interacting part of the hamiltonian corresponds to the
equivalent form of the Greens
function 
\begin{equation}
G_0^\nu=\frac{1}{z-\sum_{i=1}\frac{(V_i^\nu)^2}{z-\epsilon_i^\nu}}
\end{equation}
  which allows for an exact mapping of parameters
  $\{a_i^\nu,b_i^\nu\}\rightarrow \{V_i^\nu,\epsilon_i^\nu\}$, through
  the location 
  $G_0^\nu(\omega)|_{\omega=\epsilon_i}=0$ and derivative
  $V_i^\nu=(-d_\omega G_0^\nu|_{\epsilon_i})^{-1/2}$ of roots on the
  real axis.
\\
{\em 3) Calculate, using exact diagonalization, the interacting Greens
function $G^\nu$ and the corresponding self energy}
\begin{equation}
\Sigma^\nu(z)-\mu=(G_0^{\nu})^{-1}-(G^\nu)^{-1}\,.
\end{equation}
To be very confident of the accuracy we have
calculated $G^\nu$ as an explicit sum over all poles that correspond to
single particle excitations from the ground states,
\begin{equation}
G^\nu(z)=\frac{1}{2}\sum_{s=\uparrow,\downarrow;m}(\frac{\langle m|c^\dagger_{0,\uparrow}|0,s\rangle^2}{z-(E_m-E_0)}+\frac{\langle m|c_{0,\uparrow}|0,s\rangle^2}{z+(E_m-E_0)})
\end{equation}
 A block diagonal form in
particle number $n$, spin $S$ and $S^z$ in which the degenerate ground states are
known to be in sectors $n=n_s$, $S=1/2$ and $S^z=\pm1/2$ is used. The self energy $\Sigma^\nu(\omega)$ is then calculated by
inverting the Greens function away from the real axis at
$\omega+i\eta$.  The latter with the exception of the pole at
$\omega=0$ that we calculate explicitly as $\alpha^\nu=-\frac{1}{\partial\omega
  G^{\nu}|_{\omega=0}}$. (Removing numerically the corresponding contribution in
the inverted Greens function.) At this point the self-energy can also be
readily evaluated at any point $z$ away from the real axis for
comparison with QMC results on Matsubara frequencies.\\
{\em 4) Calculate the self energy as}
\begin{equation}
\Sigma(\omega)=\langle \Sigma^\nu(\omega)\rangle\equiv\frac{1}{N}\sum_{\nu=1}^{N}\Sigma^{\nu}(\omega)\,,
\end{equation}
with
the pole at $\omega=0$ given explicitly by
$\alpha=\frac{1}{N}\sum_\nu\alpha^\nu$. As emphasized in
Ref. \onlinecite{Dist-ED} it is {\em not} appropriate to calculate the
self energy as
$\Sigma'-\mu=\langle G_0^\nu\rangle^{-1}-\langle G^\nu\rangle^{-1}$,
because the mean of the Greens functions do not satisfy a Dyson
equation. The latter implies that roots of $G_0$ are also roots of
$G$ which will not be satisfied for the means.  \\

The calculation of the self energy is the crux of the DMFT iteration,
where $\Sigma(\omega)$ is then used to calculate a new
impurity-bath Green's function $G_0(\omega)$. The pole strength 
at $\omega=0$ is given exactly by $a_0=1/(1+\frac{1}{4\alpha})$,
making it convenient (but not crucial for the results) to
keep track of this explicitly. Importantly, the broadening $\eta$ is
built into the cycle, as the starting point of the next iteration
is based on an assumed real frequency Greens function which is actually
calculated from a self energy evaluated a finite distance from the real
axis. Thus, even if at
each iteration the dependence on $\eta$ may be quite weak, the
difference will be iteratively enhanced such that the self consistent
solutions for small and large $\eta$ turn out radically different.


\subsection*{Results and discussion}

We have done calculations using $n_s=$3,5, and 7 for $U$ ranging from
close to $U_{c1}\approx 2.5$ to very large, and
$\eta$ ranging from $0.1$ down to $10^{-5}$. We have focused primarily
on $U=4$, well into the insulating region of the phase
diagram, using $n_s=$5 and $\eta=0.001$ which are the results
discussed and shown in
the figures unless stated otherwise.
Calculations have been done on a compute cluster using up to 300
kernels, sampling up
to $N=10^7$ 5-level systems which corresponds to a self energy built up of
$10^9$ poles. This large number of stochastically distributed poles allows
for $\eta$ as small as $10^{-4}$ without significant noise. We have
used a discretization $\Delta\omega=0.001$, but since the mapping
$\Sigma(\omega)\Leftrightarrow G_0(\omega)$ is point for point this
does not introduce any additional approximation (unless there is
structure on an even smaller scale) even if
$\eta<\Delta\omega$. The only consequence is that in the latter case
we need to have a very large number of samples $N$ in order to capture
the proper $N\rightarrow\infty$ value of $\Sigma$. To initialize the calculations we use an
arbitrary insulating self energy such as the expression for
an isolated site, $\Sigma(z)\sim 1/z$.

The main result of these calculations (Fig. \ref{delta001_delta01}) that the self consistent
solutions for the self energy and corresponding spectral function has
a narrow peak structure at the inner edge of the Hubbard bands. The
peak corresponds to an actual narrow band of coherent quasiparticles 
solving $\omega-\epsilon+\mu-Re\Sigma
(\omega)=0$ for a range of $\omega$ values and that are clearly distinguished from the incoherent weight contributing
to the main part of the Hubbard bands. At $U=4$ we find a
quasiparticle weight $Z\approx \frac{1}{25}$. The quasiparticle weight
does not appear to be strongly $U$ dependent, instead the
quasiparticles are defined over a decreasing fraction of the bare band
width with increasing $U$ leading to a gradual disappearance of the
peak. Eventually, for large $U$, there is no quasiparticle solution,
but only a remnant peak. There is also weaker more strongly damped (non-quasiparticle) secondary peaks that follow from the coupled oscillations of $G(\omega)$ and
$\Sigma(\omega)$ built into the DMFT solution.

A surprising property of the calculations is that solutions are very
sensitive to the pole broadening $\eta$, with solutions converged at
$\eta=0.1$ have a featureless self energy and spectral weight with no evidence of the
quasiparticles that are found at smaller $\eta$. Operationally we find
that this qualitative difference comes form the fact that the self
energy and local Greens function are coupled self consistently which
may enhance slight differences. As
exemplified in Figure \ref{DMFT_iteration}, starting a calculation with $\eta=0.001$ from a converged solution at $\eta=0.1$
there is at each DMFT iteration only a slight enhancement at the inner edge of band but that eventually develops
into the full quasiparticle peak. 

Comparing to D-DMRG results by Karski et al.\cite{Karski_DMRG} we find that the
large $\eta$ featureless solutions are in good, semi-quantitative,
agreement. The lack of quantitative agreement may be due
to the additional step of analytic continuation from finite $\eta$ to
real frequency for the D-DMRG results. 
In contrast, as discussed before, other D-DMRG results by
Nishimoto et al. do find a peak structure close to $U_{c1}$.  


\begin{figure}
\includegraphics[scale=.5]{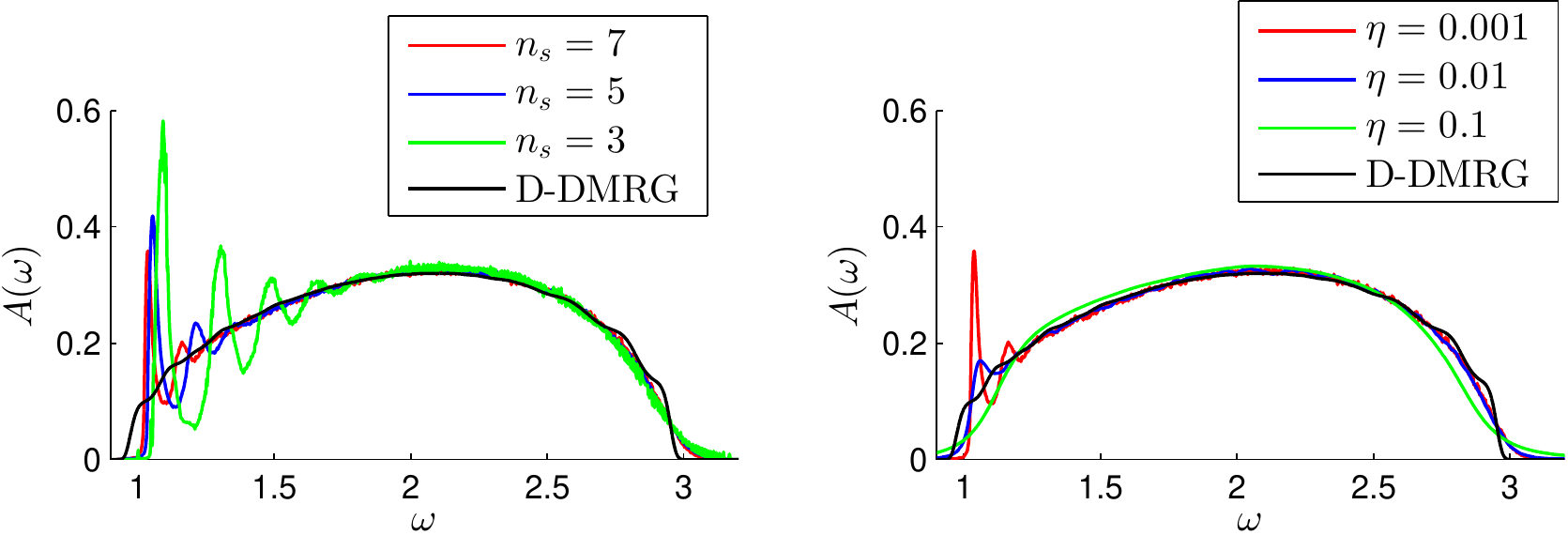}
\caption{\label{delta01_DMRG}
Uppper Hubbard band $A(\omega)$, Dist-ED converged at
various $n_s$ (with $\eta=0.001$) and various $\eta$ (with $n_s=7$) at $U=4$ and compared
to results using D-DMRG.\cite{Karski_DMRG}  
}
\end{figure}

\begin{figure}
\includegraphics[scale=.5]{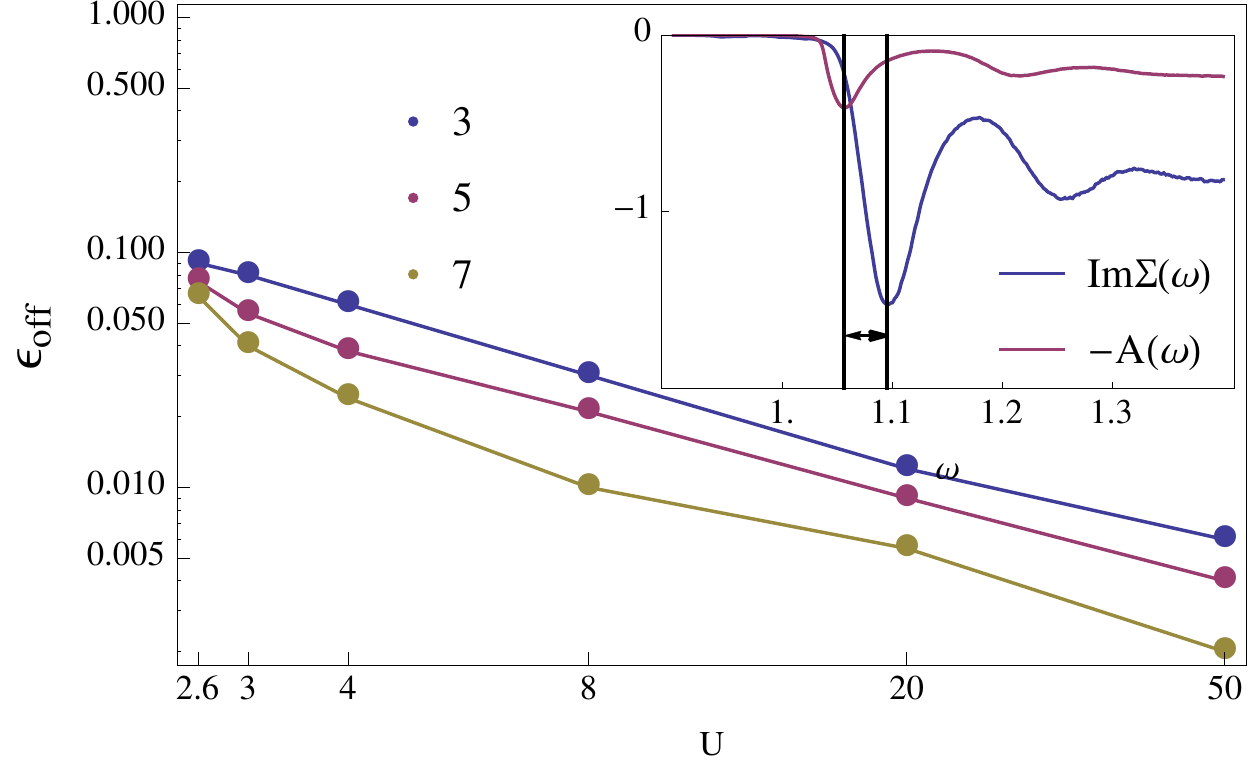}
\caption{\label{e_scale}
Energy scale $\epsilon_{\text{off}}$ of separation between the peak in $A(\omega)$ and
$Im\Sigma(\omega)$, and dependence on $n_s$.  This energy scale must
be well resolved in order to observe the quasiparticles.  (The numbers are estimates with accuracy limited primarily by the broadness of the peaks.)
}
\end{figure}

At a more basic level we find that we can identify an energy scale
that must be resolved in order to find the quasiparticle
solution. The quasiparticles that give a peak in $A(\omega)$ are caused by a sharp oscillation in the
$Re\Sigma$ which by analyticity precedes (in $|\omega|$) a peak in
$Im\Sigma$. This implies that the self-consistent solution must have
an offset $\epsilon_{\text{off}}$ between the two peaks
(see inset of Fig. \ref{e_scale}). This is exactly the outcome of the
calculation of the self energy, there is a small $U$-dependent offset
$\epsilon_{\text{off}}\leq 0.1$ between the inner edge of spectral weight (the spectral gap) and the
onset of scattering. We observe from solving finite systems, that there is such
an offset, poles in $\Sigma^{\nu}$ (identified as zeroes in $G^{\nu}$) are always
at higher energies than multiples of the bare pole locations of
$G_0^{\nu}$, an effect which seems genuinely non-perturbative. (In
perturbation theory we would expect poles in $\Sigma^{\nu}$ at
integer multiples of the poles in the bare greens function
$G_0^{\nu}$.) The crux of the matter is of course whether this offset
survives the limit $n_s\rightarrow\infty$. As shown in Figure
\ref{e_scale} the energy scale is diminished with $n_s$, although at
the smallest $U$ studied we actually find very little difference between $n_s=$3, 5, and 7.    

To put this in context, for the metallic solution (not studied here) there is an energy scale related to the
width of the central peak in $A(\omega)$ which is actually the Kondo
scale, $T_K$, of the Anderson
impurity.\cite{NRG, George_review} The energy scale identified here for the
insulator is also related to the width of a (gapped) quasiparticle
peak but whether the two energy scales are related in some way remains to be
explored. For the metal, $T_K$ vanishes at the metal-insulator
transition $U_c\approx 3$, whereas the energy scale $\epsilon_{\text{off}}$
appears to vanish only asymptotically for large $U$.


Quantum Monte Carlo calculations for quantum impurities are in principle 
numerically exact in the limit of large simulation time.\cite{QMC,CTQMC}  
Nevertheless, there is at least two major drawbacks of the method which for
the Mott insulator studied here makes the method less relevant. 
First, it works at finite temperature $\beta^{-1}=k_BT>0$, such that
when the gap scale is small there will be appreciable deviations from the $T=0$
results. 



Secondly, and more dramatic, because it is an imaginary time formalism it gives information about the Greens function or self
energy on corresponding imaginary (Matsubara) frequencies
$i\omega_n=i\frac{\pi}{\beta}(2n+1)$. To get real frequency
information requires an analytic continuation, but with statistical
noise for finite simulation times it is not feasible to study
structure at high energies on the fine scale discussed in this
paper. Nevertheless, a direct comparison on the Matsubara frequencies of the Dist-ED results
(available for any complex frequency) 
should provide a useful and rigorous test. We have used the TRIQS code
\cite{triqs} and the hybridization expansion routine within continuous
time quantum Monte Carlo (CT-QMC).\cite{CTQMC}  Within the limits set by reasonably achievable 
noise levels, $\sigma_{\text{QMC}}\approx 10^{-4}$, of the CT-QMC we find that 
the Dist-ED results are in very good agreement with the QMC for all
parameter values, which is remarkable given how different the
solutions appear over real frequencies (Fig. \ref{delta01_DMRG}). 
Only for the large $\eta=0.1$ solution (featureless on the real axis)
does there appear to be
a statistically relevant deviation from the QMC data, which may be
related to the slightly wider tails of the spectra in this
case. Clearly, the gap edge quasiparticles are consistent with QMC on the imaginary
axis, but the latter is too insensitive to the details of real
frequency structure to provide a very stringent test.

To conclude, we have found evidence that the Mott insulting state of the
Hubbard model in infinite dimensions with a semicircular density of
states has well defined gapped electronic quasiparticles
even in the paramagnetic phase. 
We have used the method of
distributional exact diagonalization\cite{Dist-ED} in which the impurity self
energy of the DMFT impurity model is
calculating as a sample average of the exact self
energy of a representative distribution of impurities models with
$n_s$ levels. Our results suggest that it is crucial to have a very good resolution at each
iteration of the DMFT cycle to resolve a small non-perturbatively
generated energy scale. 
We speculate that the quantitative details of the results presented here
will depend on the system size $n_s$, and our results for $n_s=5$ may
exaggerate the peak structure, but that the qualitative features are
genuine. This is supported by comparison to quantum Monte Carlo
results, the lack of any evident signature of finite size structure in our
calculations, as well as the earlier findings of a similar peak
feature in an independent study.\cite{Nishimoto} Nevertheless, it is
desirable to study larger system sizes within the present method to
do a systematic finite size analysis. We hope that our results will also motivate
the development of methods that can achieve similarly high resolution at
high energies to investigate this issue further. \\

We acknowledge informative discussions with Andrew Mitchell and valuable 
support from Hugo Strand and Stellan \"Ostlund. We also thank Carsten
Raas for providing the D-DMRG data from Karski et al.\cite{Karski_DMRG}.
Support was provided by the Swedish Research Council (grant
no. 2011-4054).
Computations used resources at Chalmers Centre for Computational Science and Engineering (C3SE)
provided by the Swedish National Infrastructure for Computing (SNIC).

\appendix*
\subsection{Appendix: Perturbative motivation at large $U$}

Here we demonstrate that the Dist-ED formalism applied to the
Mott insulator may be motivated as a perturbation expansion in the small 
parameter $b_0=1-a_0$, where $a_0$ is the weight of the $\omega=0$ pole of
the impurity-bath Greens function. To 0'th and 1'st order in $b_0$, the method
is exact to all orders in the standard perturbation expansion in $U$.  

Consider an arbitrary $n$'th
  order diagram in the expansion of the self-energy that contains $k=2n-1$ legs given by 
  $G_0(z)=\frac{a_0}{z}+g_0(z)$, where the calculation consists of
  evaluating Matsubara sums of products of $G_0(i\omega_n)$ (or
  correspondingly real or imaginary frequency integrals.) Suppressing
  summations we can write a diagram schematically as
  $U^n(G_0)^k=U^n(a_0+b_0)^k$ and expand in powers of $b_0$ (keeping in mind
  the underlying structure of summations that implies that the factors
  are not equivalent). We now want to study what approximation is made
  by calculating $\Sigma$ as the sample average
  $\langle \Sigma^\nu\rangle=\frac{1}{N}\sum_{\nu=1}^N\Sigma_{\nu}$ where $\Sigma^{\nu}$ is the
  exact self energy of the finite size impurity problem. For simplicity we will consider the minimal 3-level
  system with impurity-bath Green's function
  $G_0^\nu=\frac{a_0}{z}+b_0(\frac{1/2}{z-b^{\nu}}+\frac{1/2}{z+b^\nu})$,
  where the pole locations $b^\nu$ are distributed according to 
  $-\frac{1}{b_0\pi}Im g_0(w)$ such that $\langle G_0^{\nu}\rangle=\lim\limits_{N\rightarrow\infty}\frac{1}{N}\sum_{\nu=1}^NG_0^\nu=G_0$. 

The $(a_0)^k(b_0)^0$ term of the expansion corresponds to evaluating all the
Green's function in terms of the $z=0$ pole contribution. In the large
$U$ limit $a_0\rightarrow 1$ and $b\rightarrow 0$ this corresponds to
the Hubbard I approximation of an isolated site which is exact in this
limit and which is obviously exactly represented by a finite
system. To next order $(a_0)^{k-1}(b_0)^1$ we will use the second
order diagram $\Sigma^{(2)}=U^2\sum_{p,m}G_0(i\omega_{n-m})
G_0(i\omega_p) G_0(i\omega_{m-p})$ ($\omega_n=\frac{\pi}{\beta}(2n+1)$) for demonstration purpose but the proof would be
analogous for an arbitrary self energy diagram. Thus evaluating terms
proportional to $(a_0)^2b_0$ we find

\begin{widetext}
\begin{eqnarray}
\Sigma^{(2)}|_{a_0^2b_0}/U^2&=&\sum_{p,m} G_0(i\omega_{n-m}) G_0(i\omega_p) G_0(i\omega_{m-p})|_{a_0^2b_0}\nonumber
\\
&=&\sum_{p,m}  (\frac{a_0}{i\omega_{n-m}}+g_0(i\omega_{n-m}))
(\frac{a_0}{i\omega_{p}}+g_0(i\omega_p))(\frac{a_0}{i\omega_{m-p}}+g_0(i\omega_{m-p})
)|_{a_0^2b_0} \nonumber \\
&=&\sum_{p,m}(\frac{a_0}{i\omega_{n-m}}\frac{a_0}{i\omega_{p}}g_0(i\omega_{m-p})+\text{permutations
  of frequencies})\,.
\end{eqnarray}
This is to be compared to the corresponding expression in terms of a
three level system,

\begin{eqnarray}
\Sigma^{(2)}_{\text{Dist-ED}}|_{a_0^2b_0}/U^2&=&\langle\sum_{p,m}
G_0^\nu(i\omega_{n-m})G_0^\nu(i\omega_p)G_0^\nu(i\omega_{m-p})\rangle |_{a_0^2b_0}\nonumber
\\
&=&\langle\sum_{p,m}\left(\frac{a_0}{i\omega_{n-m}}+\frac{b_0}{2}(\frac{1}{i\omega_{n-m}-b^\nu}+\frac{1}{i\omega_{n-m}+b^\nu})\right)\left(\frac{a_0}{i\omega_{p}}
 +..\right)\left(\frac{a_0}{i\omega_{m-p}}+..\right)\rangle
|_{a_0^2b_0}\nonumber\\
&=&\sum_{p,m} \left(\left(\frac{a_0}{i\omega_{n-m}}\frac{a_0}{i\omega_{p}}
 \langle\frac{b_0}{2}(\frac{1}{i\omega_{m-p}-b^\nu}+\frac{1}{i\omega_{m-p}+b^\nu})\rangle\right)+\text{permutations
  of frequencies}\right)\nonumber\\
&=&\sum_{p,m}(\frac{a_0}{i\omega_{n-m}}\frac{a_0}{i\omega_{p}}g_0(i\omega_{m-p})+\text{permutations
  of frequencies})=\Sigma^{(2)}|_{a_0^2b_0}/U^2\,,
\end{eqnarray}
\end{widetext}
where we have used the fact that $b^\nu$ are
distributed according to $-Img_0$, which (using particle-hole
symmetry) implies $\langle \frac{b_0}{z\pm b^\nu}\rangle=g_0(z)$.  The
demonstrated 
calculation clearly holds for any diagram implying that the Dist-Ed
formalism is exact to 1'st order in $b_0$. Nevertheless, for large $U$ the self energy has pole
strength $\alpha\sim U^2$,\cite{strong_coupling} giving
$b_0\sim 1/U^2$, such that regarded as expansion in $1/U$ the formalism is in
fact only {\em exact} to order
$1/U^2$, with higher order terms to all orders included systematically
but approximately through the sample averaged exact diagonalization.

Going beyond first order, to order $(b_0)^2$, the representation
in terms of finite systems will not be exact. It is clear, as
expected, that larger finite systems will give better approximations.
In fact, for a flat distribution ($Im g_0(\omega)=const.$ in some
interval) we expect that a
representation in terms of five (or more) level systems will be
exact even to order $(b_0)^2$ because the calculation will give an unbiased
sampling of two energies in the support of $g_0$. The details of this
however remain to be studied in
greater depth.  

The separation of the impurity-bath Greens function $G_0$ in a pole at
$\omega=0$ with a
large fraction of the spectral weight and a gapped continuum is
special to the insulator and not valid for the
metallic solution. It is thus less clear-cut
how to best sample the continuous Greens function in terms of finite
systems. Although the Dist-ED method has been used
successfully also for the metallic problem\cite{Dist-ED} the original
work used a rather ad-hoc method of discarding samples that overrepresented the low-energy spectral
weight. How to best formulate the method for problems with continuous
low-energy weight is still under investigation.

\end{document}